\documentstyle[prl,aps]{revtex}

\begin{document}

\twocolumn
\draft{}
\bibliographystyle{try}
\topmargin 0.1cm

\newcounter{univ_counter}
\setcounter{univ_counter} {0}
\addtocounter{univ_counter} {1} 
\edef\INFNGE{$^{\arabic{univ_counter}}$ } \addtocounter{univ_counter} {1} 
\edef\JLAB{$^{\arabic{univ_counter}}$ } \addtocounter{univ_counter} {1} \edef\ODU{$^{\arabic{univ_counter}}$ } \addtocounter{univ_counter} {1} \edef\VIRGINIA{$^{\arabic{univ_counter}}$ } \addtocounter{univ_counter} {1} 
\edef\DUKE{$^{\arabic{univ_counter}}$ } \addtocounter{univ_counter} {1}  \edef\AMERICAN{$^{\arabic{univ_counter}}$ } \addtocounter{univ_counter} {1}
\edef\ASU{$^{\arabic{univ_counter}}$ } \addtocounter{univ_counter} {1} \edef\UCLA{$^{\arabic{univ_counter}}$ } \addtocounter{univ_counter} {1} \edef\CMU{$^{\arabic{univ_counter}}$ } \addtocounter{univ_counter} {1} \edef\CUA{$^{\arabic{univ_counter}}$ } \addtocounter{univ_counter} {1} \edef\SACLAY{$^{\arabic{univ_counter}}$ } \addtocounter{univ_counter} {1} \edef\CNU{$^{\arabic{univ_counter}}$ } \addtocounter{univ_counter} {1} \edef\UCONN{$^{\arabic{univ_counter}}$ } \addtocounter{univ_counter} {1} \edef\EDINBURGH{$^{\arabic{univ_counter}}$ } \addtocounter{univ_counter} {1} \edef\FIU{$^{\arabic{univ_counter}}$ } \addtocounter{univ_counter} {1} \edef\FSU{$^{\arabic{univ_counter}}$ } \addtocounter{univ_counter} {1} \edef\GWU{$^{\arabic{univ_counter}}$ } \addtocounter{univ_counter} {1} 
\edef\GLASGOW{$^{\arabic{univ_counter}}$ } \addtocounter{univ_counter} {1} \edef\INFNFR{$^{\arabic{univ_counter}}$ } \addtocounter{univ_counter} {1} \edef\ORSAY{$^{\arabic{univ_counter}}$ } \addtocounter{univ_counter} {1} \edef\ITEP{$^{\arabic{univ_counter}}$ } \addtocounter{univ_counter} {1} \edef\JMU{$^{\arabic{univ_counter}}$ } \addtocounter{univ_counter} {1} \edef\KYUNGPOOK{$^{\arabic{univ_counter}}$ } \addtocounter{univ_counter} {1} \edef\MIT{$^{\arabic{univ_counter}}$ } \addtocounter{univ_counter} {1} \edef\UMASS{$^{\arabic{univ_counter}}$ } \addtocounter{univ_counter} {1} \edef\MOSCA{$^{\arabic{univ_counter}}$ } \addtocounter{univ_counter} {1} \edef\UNH{$^{\arabic{univ_counter}}$ } \addtocounter{univ_counter} {1} \edef\NSU{$^{\arabic{univ_counter}}$ } \addtocounter{univ_counter} {1} \edef\OHIOU{$^{\arabic{univ_counter}}$ } \addtocounter{univ_counter} {1} \edef\PITT{$^{\arabic{univ_counter}}$ } \addtocounter{univ_counter} {1} \edef\RPI{$^{\arabic{univ_counter}}$ } \addtocounter{univ_counter} {1} \edef\RICE{$^{\arabic{univ_counter}}$ } \addtocounter{univ_counter} {1} \edef\URICH{$^{\arabic{univ_counter}}$ } \addtocounter{univ_counter} {1} \edef\SCAROLINA{$^{\arabic{univ_counter}}$ } \addtocounter{univ_counter} {1} \edef\UTEP{$^{\arabic{univ_counter}}$ } \addtocounter{univ_counter} {1} \edef\VT{$^{\arabic{univ_counter}}$ } \addtocounter{univ_counter} {1} \edef\WM{$^{\arabic{univ_counter}}$ } \addtocounter{univ_counter} {1} \edef\YEREVAN{$^{\arabic{univ_counter}}$ }  
\title{First Measurement of the Double Spin Asymmetry in $\vec{e}\vec{p}\rightarrow e' \pi^+ n$ in the Resonance Region\\
}
\author{R.~De~Vita,\INFNGE M.~Anghinolfi,\INFNGE V.D.~Burkert,\JLAB G.E.~Dodge,\ODU R.~Minehart,\VIRGINIA M.~Taiuti,\INFNGE H.~Weller,\DUKE G.~Adams,\RPI M.J.~Amaryan,\YEREVAN E.~Anciant,\SACLAY
% ...
\\
(The CLAS Collaboration)} 
\address{\INFNGE Istituto Nazionale di Fisica Nucleare, Sezione di Genova, and Dipartimento di Fisica dell'Universit\`a, 16146 Genova, Italy}
\address{\JLAB Thomas Jefferson National Accelerator Laboratory, Newport News, Virginia 23606, USA}
\address{\ODU Old Dominion University, Norfolk, Virginia 23529, USA}
\address{\VIRGINIA University of Virginia, Charlottesville, Virginia 22901, USA}
\address{\DUKE Duke University, Durham, North Carolina 27708-0305, USA}
\address{\AMERICAN American University, Washington, D.C. 20016, USA}
\address{\ASU Arizona State University, Tempe, Arizona 85287-1504, USA}
\address{\UCLA University of California at Los Angeles, Los Angeles, California 90095-1547, USA}
\address{\CMU Carnegie Mellon University, Pittsburgh, Pennslyvania 15213, USA}
\address{\CUA Catholic University of America, Washington, D.C. 20064, USA}
\address{\SACLAY CEA-Saclay, Service de Physique Nucl\'eaire, F91191 Gif-sur-Yvette, Cedex, France}
\address{\CNU Christopher Newport University, Newport News, Virginia 23606, USA}
\address{\UCONN University of Connecticut, Storrs, Connecticut 06269, USA}
\address{\EDINBURGH Edinburgh University, Edinburgh EH9 3JZ, United Kingdom}
\address{\FIU Florida International University, Miami, Florida 33199, USA}
\address{\FSU Florida State University, Tallahasee, Florida 32306, USA}
\address{\GWU The George Washington University, Washington, DC 20052}
\address{\GLASGOW University of Glasgow, Glasgow G12 8QQ, United Kingdom} 
\address{\INFNFR Istituto Nazionale di Fisica Nucleare, Laboratori Nazionali di Frascati, P.O. 13, 00044 Frascati, Italy}
\address{\ORSAY Institut de Physique Nucleaire ORSAY, Orsay, France}
\address{\ITEP Institute of Theoretical and Experimental Physics, Moscow, 117259, Russia}
\address{\JMU James Madison University, Harrisonburg, Virginia 22807, USA}
\address{\KYUNGPOOK Kyungpook National University, Taegu 702-701, South Korea}
\address{\MIT Massachusetts Institute of Technology, Cambridge, Massachusetts  02139-4307, USA}
\address{\UMASS University of Massachusetts, Amherst, Massachusetts 01003, USA}
\address{\MOSCA Moscow State University, Moscow, 119899 Russia}
\maketitle
\address{\UNH University of New Hampshire, Durham, New Hampshirs 03824-3568, USA}
\address{\NSU Norfolk State University, Norfolk, Virginia 23504, USA}
\address{\OHIOU Ohio University, Athens, Ohio 45701, USA}
\address{\PITT University of Pittsburgh, Pittsburgh, Pennslyvania 15260, USA}
\address{\RPI Rensselaer Polytechnic Institute, Troy, New York 12180-3590, USA}
\address{\RICE Rice University, Houston, Texas 77005-1892, USA}
\address{\URICH University of Richmond, Richmond, Virginia 23173, USA}
\address{\SCAROLINA University of South Carolina, Columbia, South Carolina 29208, USA}
\address{\UTEP University of Texas at El Paso, El Paso, Texas 79968, USA}
\address{\VT Virginia Polytechnic Institute and State University, Blacksburg, Virginia 24061-0435, USA}
\address{\WM College of Willliam and Mary, Williamsburg, Virginia 23187-8795, USA}
\address{\YEREVAN Yerevan Physics Institute, 375036 Yerevan, Armenia}
\widetext
\begin{center}
$^{27}${\it University of New Hampshire, Durham, New Hampshirs 03824-3568, USA}\\
$^{28}${\it Norfolk State University, Norfolk, Virginia 23504, USA}\\
$^{29}${\it Ohio University, Athens, Ohio 45701, USA}\\
$^{30}${\it University of Pittsburgh, Pittsburgh, Pennslyvania 15260, USA}\\
$^{31}${\it Rensselaer Polytechnic Institute, Troy, New York 12180-3590, USA}\\
$^{32}${\it Rice University, Houston, Texas 77005-1892, USA}\\
$^{33}${\it University of Richmond, Richmond, Virginia 23173, USA}\\
$^{34}${\it University of South Carolina, Columbia, South Carolina 29208, USA}\\
$^{35}${\it University of Texas at El Paso, El Paso, Texas 79968, USA}\\
$^{36}${\it Virginia Polytechnic Institute and State University, Blacksburg, Virginia 24061-0435, USA}\\
$^{37}${\it College of Willliam and Mary, Williamsburg, Virginia 23187-8795, USA}\\
$^{38}${\it Yerevan Physics Institute, 375036 Yerevan, Armenia}\\
\date{February 7, 2001}
\end{center}
\maketitle

\newpage

\wideabs{
\begin{abstract}
The double spin asymmetry in the $\vec{e}\vec{p}\rightarrow e' \pi^+ n$ reaction has been measured for the first time in the resonance region for four-momentum transfer $Q^2=0.35-1.5$ GeV$^2$. Data were taken at Jefferson Lab with the CLAS detector using a 2.6 GeV polarized electron beam incident on a polarized solid NH$_3$ target. Comparison with predictions of phenomenological models shows strong sensitivity to resonance contributions. Helicity-1/2 transitions are found to be dominant in the second and third resonance regions. The measured asymmetry is consistent with a faster rise with $Q^2$ of the helicity asymmetry $A_1$ for the $F_{15}(1680)$ resonance than expected from the analysis of the unpolarized data.
\end{abstract}
}

\narrowtext

Measurements of the spin structure of exclusive pion production provide a new approach to an understanding of the structure of baryon resonances, which has been the subject of experimental and theoretical studies for many years \cite{stoler}. The nucleon and its resonant states can, in principle, be described by QCD in terms of their elementary constituents, {\it i.e.} quarks and gluons.
The complex, non-perturbative nature of resonance transitions makes this a difficult task, not fully accessible with the currently employed techniques of QCD. Therefore models continue to play an important role in the interpretation of the data, and in attempts of obtaining a better understanding of the effective degrees underlying nucleon structure in the regime of confinement. Electroexcitation of resonances is a powerful tool to address these issues. In electroproduction, the transition to resonant states is characterized by the transverse helicity amplitudes $A_{1/2}$ and $A_{3/2}$, and by the longitudinal amplitude $S_{1/2}$, where 1/2 and 3/2 refer to the total helicity of the $\gamma^*N$ system. The $Q^2$ dependence of these amplitudes yields information on the spin structure of the transition and on the wave function of the excited state. Models of baryon resonances, especially constituent quark models \cite{close,warns,capstick,giannini}, make predictions for these quantities and can be tested via comparison with the measured amplitudes. Comparison with experimental data requires proper treatment of the hadronic final state and inplementation of the non-resonant part of the pion production amplitude, which are usually not included in quark models. Phenomenological models, such as MAID \cite{maid} and AO \cite{ao}, parameterize the full amplitude through inclusion of s-channel resonances, non-resonant terms, and decay into the pion channel  from pion-nucleon scattering experiments. Such models can therefore predict the full pion electroproduction cross section and polarization asymmetries. Sensitivity to quark models predictions can be studied by varying the resonant contributions according to the models.

Single pion production has been one of the main sources of information for these studies. However, most of these experiments have been limited to the measurement of the unpolarized cross section. Only recently have technological developments in polarized sources and targets opened new possibilities for the study of polarization observables. These quantities provide important new constraints for the extraction of the resonance parameters. Double polarization experiments directly probe the helicity structure of the reaction, allowing the separation of the helicity amplitudes $A_{1/2}$ and $A_{3/2}$ without the complex analysis of the full angular distribution that is necessary for unpolarized measurements. Double spin observables in single pion photoproduction have been measured in recent experiments at MAMI \cite{mainz}, while only recoil polarization measurements have been performed in electroproduction \cite{mainz2}. In addition to a highly-polarized beam and target, or the use of recoil polarimeters, these measurements require a large acceptance detector to measure the full angular distribution of the outgoing pion and to compensate for the relatively low luminosity that polarized solid targets can tolerate.

In this letter we present the first measurement of double spin asymmetry in $\vec{e}\vec{p}\rightarrow e' \pi^+ n$, performed with the CEBAF Large Acceptance Spectrometer (CLAS) \cite{clas} in Hall B at Jefferson Lab. CLAS provides the large angular coverage that is necessary for the study of resonance decays. It is a magnetic spectrometer based on a six-coil torus magnet whose field is primarily oriented along the azimuthal direction. The particle detection system includes drift chambers for track reconstruction\cite{dc}, scintillation counters for the time of flight measurement \cite{sc}, Cerenkov counters for electron-pion discrimination \cite{cc}, and electromagnetic calorimeters to identify electrons and neutrals \cite{ec}. Charged particles can be detected and identified for momenta down to 0.2 GeV. With the polarized target inserted in the field-free region at the center of the detector, the acceptance for polar angles is restricted to the regions \mbox{$8^\circ <\theta< 50^\circ$ and  $75^\circ <\theta< 105^\circ$}. 

\begin{figure}[h]
\vspace{5.cm}
\includegraphics{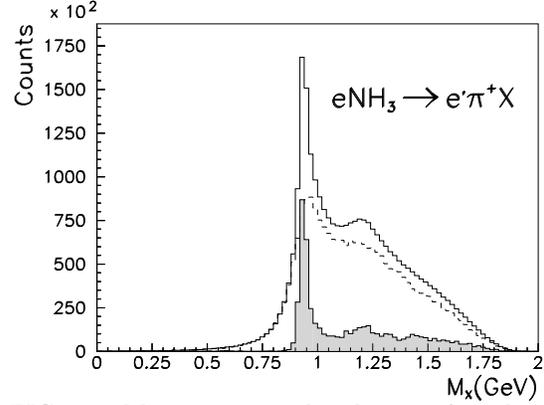}
\caption[]{Missing mass distribution for the reaction $ep\rightarrow e'\pi^+X$. The three overlapping histograms represent the $NH_{3}$ (solid line), nuclear background (dashed line), and derived hydrogen (filled histogram) spectra, respectively. The shape of the neutron peak is affected by radiative effects which are responsible of the tail on the right side of the peak.}
\label{fig:mm}
\end{figure}

Data were taken at a beam energy of 2.6 GeV. The electron beam, with an average longitudinal polarization of 70$\%$, was incident on a cylindrical 1.5 cm diameter 1 cm length target cell filled with solid NH$_3$ pellets. The beam helicity was flipped at a rate of 1 Hz in a pseudo-random sequence to minimize systematic effects. The target material was maintained at a temperature of 1 K in a 5 T magnetic field generated by a superconducting Helmholtz magnet with its axis on the beamline. A proton polarization of 50-70\%  parallel to the beam was obtained by the technique of dynamic nuclear polarization \cite{dnp}. The scattered electron was detected by a coincidence of Cerenkov counter and electromagnetic calorimeter. The positive pion was identified in coincidence with the electron by comparing its momentum, determined from the reconstructed track, and its time of flight, as measured by the scintillators. A cut on the reconstructed missing mass from the $e'\pi^+$ system of $0.85<M_x< 1.05$ GeV was used to select the exclusive $e\pi^+n$ final state \mbox{(see Figure \ref{fig:mm})}.

The cross section of $\pi^+$ electro-production with polarized beam and target can be written as,
\begin{equation}
\sigma=\sigma_0+P_e\sigma_e+P_p\sigma_p+P_eP_p\sigma_{ep},
\end{equation}
\noindent
where $\sigma_0$ is the unpolarized cross section and $P_e$ and $P_p$ refer to the electron and proton polarization, respectively. Data with different combinations of the electron and proton polarization were used to isolate the double spin term, $\sigma_{ep}$, and to extract the double spin asymmetry, defined as $A_{ep}=-\sigma_{ep}/\sigma_0$. After integrating over the pion azimuthal center-of-mass angle $\phi_{\pi}^*$, this quantity can be parameterized as \cite{bm},
\begin{equation}
A_{ep}=\sqrt{1-\epsilon^2}\cos\theta_{\gamma}\frac{A_1+\eta A_2}{1+\epsilon R},
\end{equation}
\noindent 
where $\epsilon=\left [1+2|q|^2/Q^2\tan\theta_e/2\right ]^{-1}$ is the virtual photon polarization, $q$ is the three-vector momentum transfer, $\theta_{\gamma}$ is the angle between the target spin and the virtual-photon momentum direction, \mbox{$\eta=\tan\theta_{\gamma}\sqrt{2\epsilon/(1+\epsilon)}$}, and $R$ is the longitudinal-transverse cross section ratio \mbox{$\sigma_L/\sigma_T$}. The structure function $A_1$ is the virtual photon helicity asymmetry,
\begin{equation}
A_1=\frac{|A_{1/2}|^2-|A_{3/2}|^2}{|A_{1/2}|^2+|A_{3/2}|^2},
\end{equation}
\noindent 
while $A_2$ is a longitudinal-transverse interference term. In $\pi^+$ electroproduction, the longitudinal coupling is due to the pion-pole contribution and to the resonance excitation. Analysis of unpolarized measurements, as the one of Ref. \cite{gerhardt}, and predictions of models such as MAID and AO showed that the longitudinal resonance couplings are smaller than the transverse ones. This has been confirmed in the $P_{33}(1232)$ region by the recent measurement of Ref. \cite{pizero}. The asymmetry $A_{ep}$ is therefore expected to be dominated by the purely transverse term $A_1$.

The $e'\pi^+n$ events were accumulated in bins of $Q^2$, $W$, and $\cos\theta_{\pi}^*$. To increase statistics, the data were integrated over the azimuthal angle $\phi_{\pi}^*$.  Geometrical cuts were used to select the high efficiency regions of the detector excluding the edges of the Cerenkov counter and of the drift chambers, and malfunctioning scintillators. For each event, the center-of-mass phi-averaged acceptance was analytically calculated by projecting the fiducial acceptance into the center-of-mass frame. The systematic uncertainty on the asymmetry due to the acceptance evaluation was estimated to be $\sim 0.01-0.02$. The double spin asymmetry was then obtained as,
\begin{equation}\label{eq:asy}
A_{ep}=\frac{1}{fP_eP_p}\frac{N(\uparrow\downarrow)+N(\downarrow\uparrow)-N(\uparrow\uparrow)-N(\downarrow\downarrow)}{N(\uparrow\downarrow)+N(\downarrow\uparrow)+N(\uparrow\uparrow)+N(\downarrow\downarrow)},
\end{equation}
\noindent
where $N$ represents the yields of the exclusive $e'\pi^+n$ final state, the arrows in parenthesis refer to the electron and proton spin orientation, and $f$ denotes the dilution factor for the NH$_3$ target. The asymmetry $(A_1+\eta A_2)/(1+\epsilon R)$ was extracted dividing the double spin asymmetry by the factor $\sqrt{1-\epsilon^2}\cos\theta_{\gamma}$.

The dilution factor represents the fraction of events originating from polarized-target nucleons, and accounts for the contribution of the nuclear background from the liquid helium, $^{15}$N, and vacuum windows in the target. Separate measurements on $^{12}$C and liquid helium were used to model the nuclear background distribution, shown as the dashed line in Figure \ref{fig:mm}. The background spectrum was normalized to the NH$_3$ spectrum in the tail of the missing mass peak ($M_x <0.85$ GeV) where only nuclear reactions can contribute. The systematic error on the measured asymmetry associated with this procedure was estimated to be $\sim0.04$. A subsequent measurement made directly with a solid $^{15}$N target confirmed the validity of this method used to extract the dilution factor. 

The beam and target polarizations, $P_e$ and $P_p$, were routinely monitored during data taking by a M{\o}ller polarimeter and a NMR system, respectively. A more precise value of the product $P_eP_p$ was extracted from the simultaneously measured asymmetry for elastic electron-proton scattering, which only depends on the known proton form factors and on the kinematics. Elastic events were selected by measuring the momentum and angle of the scattered electron. An independent analysis was performed with elastic events from electron-proton coincidences. The results obtained with these different elastic event selections were in excellent agreement. The error due to uncertainties in the parameterization of the form factors was estimated as 1$\%$, and the overall relative accuracy in $P_eP_p$ was $2-3\%$. 

Radiative corrections were calculated using a Monte Carlo integration of the Mo and Tsai formula \cite{motsai}. Two different models \cite{maid,ao} were used to generate the Born cross section and the discrepancy in the calculated asymmetries ($\sim0.01-0.02$) was used as an estimate of the model dependency of the correction.

Systematic effects deriving from the event selection, as for example the missing mass cut, were studied by varying the selection criteria. The total systematic error on the double spin asymmetry due to all sources discussed was estimated to be on the average $\sim0.05-0.06$, which is much smaller than the statistical error.

\begin{figure}[h]
\vspace{7.8cm}
\includegraphics{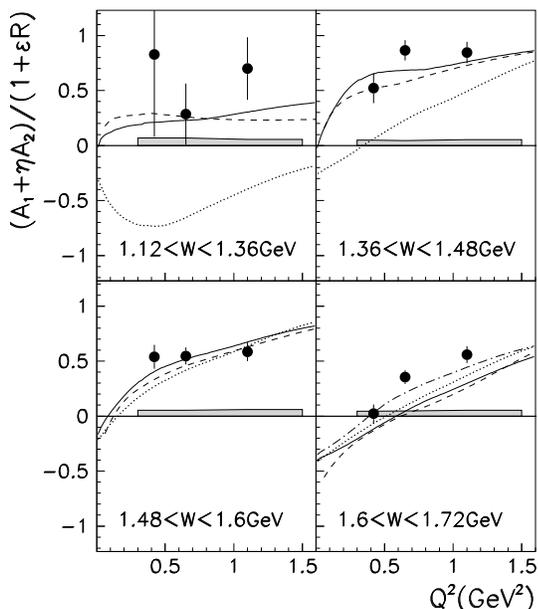}
\caption[]{$Q^2$ dependence of the double spin asymmetry $(A_1+\eta A_2)/(1+\epsilon R)$. The error bars show the statistical error while the shaded bands represent the systematic uncertainty. The data are compared with the pure resonance contribution (dotted line) predicted by the AO model \cite{ao}, with the MAID \cite{maid}(dashed line) and AO (solid line) full calculations, and with the AO prediction obtained with a $\Delta A_1 \sim 0.40$ increase for the $F_{15}(1680)$ (dashed-dotted line).}
\label{fig:q2_dep}
\end{figure}

The double spin asymmetry was evaluated in three $Q^2$ bins ranging from 0.35 to 1.5 GeV$^2$, and in three bins in the angular range $0.25 <\cos\theta_{\pi}^*< 1.0$, where the acceptance is complete. For each $Q^2$ and $\cos\theta_{\pi}^*$ bin, the $W$ dependence was measured from 1.12 GeV up to a maximum of 1.96 GeV. The results were compared to the MAID \cite{maid} and the AO \cite{ao} models. For the MAID model, the electromagnetic multipoles up to $L=5$ were used to calculate the helicity amplitudes and the resulting response functions for this process. The cross section terms were integrated over the same bins and acceptance covered by the data, in order to provide a direct comparison. A similar procedure was used for the AO calculation, starting in this case directly from the helicity amplitudes given in this program. 


The $Q^2$ dependence of the asymmetry $(A_1+\eta A_2)/(1+\epsilon R)$ integrated over $\cos\theta_{\pi}^*$ is shown in Figure \ref{fig:q2_dep} for four $W$ ranges. The dotted curve represents the pure resonance contribution as predicted by the AO model, while the solid and dashed lines are, respectively, the AO and MAID calculations including non-resonant amplitudes. In the low $W$ region, the asymmetry is strongly affected by non-resonant processes, leading to positive values in spite of the negative asymmetry expected for the $P_{33}(1232)$ state. For $W >1.48$ GeV, the resonance contribution becomes dominant and the asymmetry is positive, indicating that the reaction is ruled by the helicity-1/2 amplitude. This is in contrast with the helicity-3/2 dominance observed at the photon point \cite{pdg} and indicates that a transition occurs in between $Q^2=0$ and the measured $Q^2$ range. This feature is consistent with a strong change with $Q^2$ of the helicity structure of the $D_{13}(1520)$ and $F_{15}(1680)$ states that are predicted by constituent quark models \cite{close,warns,capstick,giannini} to vary from $A_1=-1$ at the photon point to $A_1=1$ at high $Q^2$. In the second resonance region ($1.48<W<1.6$ GeV) the asymmetry is large already at small $Q^2$, and slowly approaching saturation, while in the third resonance region ($1.6<W<1.72$ GeV) the rise with $Q^2$ indicates a slower transition from helicity-3/2 to helicity-1/2 dominance due to the underlying $F_{15}(1680)$.
\begin{figure}[h]
\vspace{7.8cm}
\includegraphics{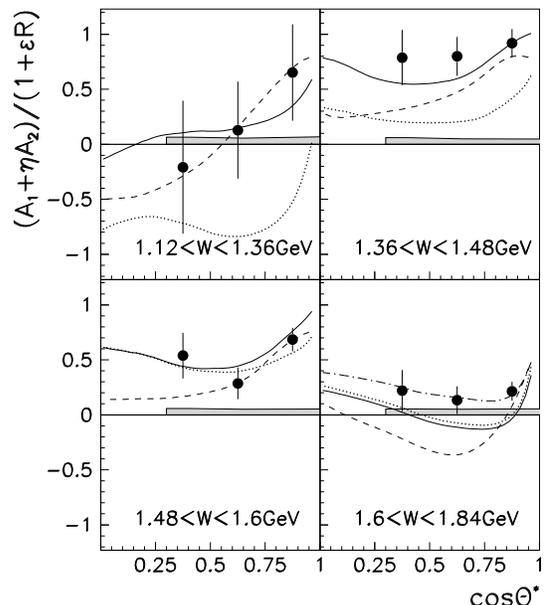}
\caption[]{Angular dependence of the double spin asymmetry $(A_1+\eta A_2)/(1+\epsilon R)$ for $0.5 < Q^2 < 0.9$ GeV$^2$. The error bars show the statistical error while the shaded bands represent the systematic uncertainty. The curves are from the same models as in Figure 2.}
\label{fig:ang_dep}
\end{figure}
Additional insight into the helicity structure of the process is provided by the study of the angular distribution. Figure \ref{fig:ang_dep} shows the angular dependence of the asymmetry $(A_1+\eta A_2)/(1+\epsilon R)$ for $0.5<Q^2<0.9$ GeV$^2$ in four different $W$ bins. The rise at forward angles is due to angular momentum conservation that constrains the helicity asymmetry to 1 at $\theta_{\pi}^*=0$. This is evident in the $P_{33}(1232)$ region where the asymmetry changes sign both because of this constraint and because of the competing contribution of the background that is dominant at forward angles and is predicted to give a positive asymmetry. In Figure \ref{fig:ang_dep} the four curves are generated in the same way as in Figure \ref{fig:q2_dep}. Both models agree fairly well with our results in the low $W$ region. At higher $W$ a systematic discrepancy between the CLAS data and the MAID prediction appears for the lower $\cos\theta_{\pi}^*$ bin, indicating that the model may underestimate the helicity-1/2 contribution in the second and third resonance regions. A better quantitative agreement is found with the AO calculations. The AO model was modified to include a new parameterization of the resonance amplitudes for the $[70,1^-]$ multiplet based on recent measurements of the photo-coupling for the $S_{11}(1535)$ resonance \cite{etab,etac} and predicts a larger helicity-1/2 amplitude in the second resonance region than previous parameterizations \cite{maid,sqtm_ao}.

A study on the sensitivity to single resonance contributions was performed for the highest $W$ interval. The discrepancy between the data and the model predictions shown both by the $Q^2$ and the angular dependence for the highest $W$ interval was found to be compatible with a $\Delta A_1 = 0.40$ increase of the $F_{15}(1680)$ helicity asymmetry included in the AO model. Similar variations applied to other excited states that contribute to this $W$ range did not result in significant improvements in the agreement of the AO calculations with the CLAS data. This result indicates that the already mentioned transition of the $F_{15}(1680)$ from helicity 3/2 to helicity 1/2 dominance may be more rapid than what originally implemented in the AO and MAID models and suggested by the unpolarized data. The new $Q^2$ dependence indicated by the CLAS data is in agreement with the prediction of the relativistic quark model of Ref. \cite{capstick}.  


In conclusion, we have presented the first measurement of the double spin asymmetry for the $\vec {e}\vec{p} \rightarrow e'\pi^+n$ channel. A comparison with phenomenological models shows the high sensitivity of this observable to resonance contributions. The sign and magnitude of the measured asymmetry indicate the dominance of the helicity-1/2 contribution in the reaction, in contrast with the helicity-3/2 dominance observed at the photon point. 
The helicity flip seen at low W is qualitatively consistent with expectations from phenomenological analyses of unpolarized data as represented in MAID and AO, while for $W >1.6$ GeV the changeover to helicity 1/2 dominance occurs at lower $Q^2$ than expected in both descriptions. This feature is in agreement with a rapid change in the helicity structure of electromagnetic transitions in the region of the $F_{15}(1680)$ resonance, predicted by the relativized constituent quark model \cite{capstick}.

We would like to acknowledge the outstanding efforts of the staff of the Accelerator, Target Group, and Physics Division at TJNAF that made this experiment possible. We are grateful to D.~Drechsel and L.~Tiator for useful discussions. This work was supported by the Italian Istituto Nazionale di Fisica Nucleare, the French Commissariat \`a l'Energie Atomique, the U.S. Department of Energy and National Science Foundation, and the Korea Science and Engineering Foundation. The Southeastern Universities Research Association (SURA) operates the Thomas Jefferson National Accelerator Facility for the United States
Department of Energy under contract DE-AC05-84ER40150.

\end{document}